\documentclass[twocolumn,9pt]{article} 

\usepackage[square,numbers,sort&compress,comma]{natbib}

\usepackage{amsmath}
\usepackage{amssymb}
\usepackage{caption}
\usepackage{graphicx}
\usepackage{latexsym}
\usepackage{times}
\usepackage[pagewise]{lineno}
\usepackage{hyperref}
\usepackage{mhchem}
\usepackage{xcolor}

\topmargin - 12pt 
\renewenvironment{abstract}%
              {
               \small
               {\bfseries \abstractname}
               \par
               \vspace{10pt}
              }

\renewcommand\abstractname{Abstract}

\newcommand{\nomenclature}
              [1]
              {
               \bgroup
               \flushleft
               \small\bf
               #1
               \par
               \egroup
              }

\renewcommand{\section}
              [1]
              {
               \bgroup
               \flushleft
               \small\bf
               \refstepcounter{section}
               \arabic{section}. #1
               \par
               \egroup
              }

\renewcommand{\subsection}
              [1]
              {
               \bgroup
               \flushleft
               \small\em
               \refstepcounter{subsection}
               \arabic{section}.
               \arabic{subsection}. #1
               \par
               \egroup
              }

\renewcommand{\subsubsection}
              [1]
              {
               \bgroup
               \flushleft
               \small\em
               \refstepcounter{subsubsection}
               \arabic{section}.
               \arabic{subsection}.
               \arabic{subsubsection}. #1
               \par
               \egroup
              }

  \newcommand{\acknowledgement}
              [1]
              {
               \bgroup
               \flushleft
               \small\bf
               #1
               \par
               \egroup
              }

  \newcommand{\sectionbib}
              [1]
              {
               \bgroup
               \flushleft
               \small\bf
               #1
               \par
               \egroup
              }

\begin{document}



\small
\baselineskip 10pt

\setcounter{page}{1}
\title{\LARGE \bf Effects of Soret diffusion on the intrinsic instability of premixed hydrogen/air flames}

\author{{\large Qizhe Wen$^{a}$, Yan Wang$^{a}$, Linlin Yang$^{a}$, Youhi Morii$^{b}$,  Thorsten Zirwes$^{c}$},\\ {\large Shengkai Wang$^{a*}$, Zheng Chen$^{a}$}\\[10pt]
        {\footnotesize \em $^a$SKLTCS, HEDPS, School of Mechanics and Engineering Science, Peking University, Beijing, 100871, P.R. China}\\[-5pt]
        {\footnotesize \em $^b$Institute of Fluid Science, Tohoku University, 2-1-1 Katahira, Aoba-ku, Sendai, 9808577, Japan}\\[-5pt]
        {\footnotesize \em $^c$ Institute for Reactive Flows (IRST), University of Stuttgart, Pfaffenwaldring 31, Stuttgart, 70569, Germany}}

\date{}  

\twocolumn[\begin{@twocolumnfalse}
\maketitle
\rule{\textwidth}{0.5pt}
\vspace{-5pt}

\begin{abstract} 
Hydrogen flames exhibit multiple intrinsic instabilities. The low molar masses of \ce{H} and \ce{H2} lead to significant Soret diffusion near the flame front; however, its influence on hydrogen flame instabilities remains to be quantified. This study investigates the effect of Soret diffusion on instability evolution dynamics via one-dimensional counterflow analysis and two-dimensional, high-fidelity direct numerical simulations covering both the linear growth regime and the fully developed nonlinear regime over a wide range of equivalence ratios ($\phi$). In the linear regime, Soret diffusion increases the perturbation growth rate at $\phi < 1.7$, especially under lean conditions, but reduces the growth rate at $\phi > 1.7$. A similar sensitivity reversal is observed in the Markstein length ($\mathcal{L}$) across the critical equivalence ratio $\phi_c = 1.7$, which coincides with the peak equivalence ratio of unstretched laminar flame speed. In the nonlinear regime, Soret diffusion accelerates the formation of small-scale wrinkles in lean hydrogen flames and reduces the characteristic size of large-scale finger-like structure by one-third. An interesting observation is that, although Soret diffusion promotes preferential diffusion and increases the local flame displacement speed, the global fuel consumption rate decreases due to a substantial reduction in the overall flame surface area. In addition, curvature-based flame segment analysis reveals a synergistic effect between Soret diffusion and Fickian diffusion that enhances/reduces the local equivalence ratio in positively/negatively curved regions of the flame front. The probability distributions of the Karlovitz number ($Ka$) and the density-weighted displacement speed ($S^*_d$) are also analyzed; results suggest that, for lean hydrogen flames, Soret diffusion broadens the distributions for both parameters, particularly on the positive side. These findings promise to improve the understanding of hydrogen flame dynamics under complex differential transport across an extended range of equivalence ratios and through different regimes of instability evolution.

\end{abstract}

\vspace{10pt}

{\bf Novelty and Significance Statement}
\vspace{10pt}

This study addresses an important gap regarding the effect of Soret diffusion on hydrogen flame instability, particularly in the nonlinear regime of flame evolution. It extends beyond conventional linear stability theory and quantifies how Soret diffusion influences the probability distributions of key properties (e.g., curvature, flame speed, local equivalence ratio, and species production rates) at long times when they approach a quasi-steady state. Moreover, unlike prior work that primarily focused on lean hydrogen flames, the present study examines a broader range of equivalence ratios and identifies a critical equivalence ratio at which the Soret-diffusion sensitivity of flame instability reverses. A counterintuitive finding is that Soret diffusion can simultaneously increase local flame displacement speed and reduce the global fuel consumption rate. The underlying mechanisms are analyzed, and a physical link is established between molecular cross-diffusion, multi-scale morphology evolution, and flame propagation. The results should improve the understanding of hydrogen flame dynamics.

\vspace{5pt}
\parbox{1.0\textwidth}{\footnotesize {\em Keywords:} Soret effect; Hydrogen; Flame instability; Flame Speed; Premixed flame }
\rule{\textwidth}{0.5pt}
*Corresponding author.
\vspace{5pt}
\end{@twocolumnfalse}] 

\section{Introduction}\label{sec:intro}\addvspace{10pt}

Hydrogen, a zero-carbon fuel, has attracted significant attention as a clean and sustainable alternative to fossil fuels~\cite{mallapaty2020china}. However, due to the exceptionally high mass diffusivity of hydrogen molecules, hydrogen flames exhibit a variety of intrinsic instability mechanisms. Among these mechanisms, the Darrieus–Landau (DL) instability~\cite{landau1988theory, darrieus1938propagation}, caused by thermal expansion or the density difference across the flame front, is widely observed in premixed flames. Meanwhile, the thermodiffusive (TD) instability~\cite{manton1952nonisotropic, markstein2014nonsteady, palm1969propagation}, which arises from the coupling between flame stretch and differential transport of mass and heat, is particularly pronounced in lean hydrogen flames. These instabilities often lead to the formation of complex cellular structures along the flame front, significantly affecting flame propagation and combustion characteristics. Consequently, a comprehensive understanding of hydrogen flame instabilities is essential for the safe and efficient utilization of hydrogen as an energy carrier. Over the past few years, extensive theoretical \cite{sivashinsky1977diffusional,matalon1982flames,Sivashinsky1983Instabilities,clavin1985dynamic,matalon2003hydrodynamic, law2010combustion} and numerical \cite{berger2019characteristic, berger2022intrinsic_a, gaucherand2023intrinsic, frouzakis2015numerical, altantzis2012hydrodynamic, yang2025propagation} studies have been conducted to elucidate their underlying mechanisms. 

Notably, the extremely low molecular mass of hydrogen renders its flames highly susceptible to Soret diffusion -- a mass transport phenomenon driven by temperature gradients. The impact of Soret diffusion on flame speeds has been widely studied for hydrogen at atmospheric~\cite{ern1998thermal, ern1999impact, yang2010mechanistic} and elevated pressures~\cite{faghih2018effects}, as well as for heavy hydrocarbons~\cite{yang2011effects,xin2012mechanistic} and syngas~\cite{liang2013effects}. Its effect on ignition~\cite{garcia1994analysis, han2015initiation, jayachandran2017thermal, gopalakrishnan2004effects, 2024YuChenIgnit} and extinction~\cite{fong2012asymptotic, han2015counterflow, yang2010mechanistic, xin2012mechanistic, ern1999impact} dynamics has also been extensively explored theoretically and numerically. Specifically, Liang et al.~\cite{liang2018ignition} found that Soret diffusion decreases the Markstein length at low equivalence ratios but increases it at high equivalence ratios, though the exact transition point was not reported. In addition, the influence of Soret diffusion on pollutant formation~\cite{rosner2000heavy, dworkin2009impact, acquaviva2025influence} has been extensively investigated.

Regarding the effect of Soret diffusion on flame instability, early asymptotic analysis by Garcia-Ybarra et al.~\cite{garcia1984soret} demonstrated that Soret diffusion significantly alters the TD stability limits of premixed flames. In recent numerical studies, Grcar et al.~\cite{grcar2009soret} found that Soret diffusion increases the local fuel consumption rate of cellular flames and promotes earlier cell division. Zhou et al. \cite{zhou2017effect} also noted that Soret diffusion affects the early stage of cellular flame propagation. D'Alessio et al.~\cite{d2024intrinsic} reported that Soret diffusion increases the maximum linear growth rate and the cutoff wavenumber in lean \ce{H2}/\ce{NH3}/Air flames. Zirwes et al.~\cite{zirwes2024role} found that Soret diffusion enriches the local mixture in the reaction and post-oxidation zones of lean \ce{H2} flames, thereby increasing their heat release rates and propagation speeds. Howarth et al. proposed a stretch factor model for 2D cellular flames ~\cite{howarth2022empirical} and confirmed that it remains applicable when Soret diffusion is included ~\cite{howarth2024thermal}. 

Despite progress to date, previous studies have focused primarily on lean conditions, and investigations of rich mixtures have been largely confined to pulsating instabilities ~\cite{korsakova2016effect}. Across a broader range of equivalence ratios, the underlying physical mechanisms by which Soret diffusion affects flame instabilities remains to be fully elucidated. In addition, the influence of Soret diffusion on flame morphology characteristics in the nonlinear regime of instability evolution, especially at long times where the probability distributions of key flame properties approach a quasi-steady state, remains to be quantified.

To address these issues, high-fidelity numerical simulations of \ce{H2}/air flames spanning lean to rich conditions are conducted. Both the linear and nonlinear regimes are analyzed to elucidate the underlying mechanisms.

\section{Method}\label{sec:Method}\addvspace{10pt}

\begin{figure}
    \centering
    \includegraphics[width=0.75\linewidth]{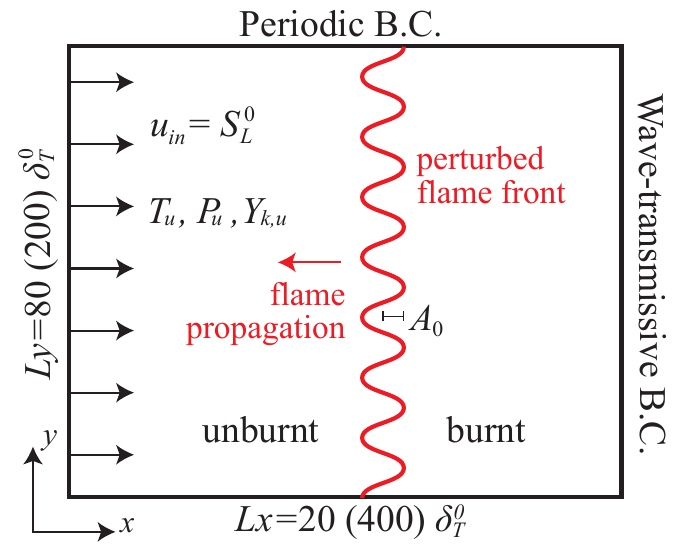}
    \caption{Computational setup and boundary conditions. The dimensions of the computation domain for linear (nonlinear) analysis are specified by $L_x$ and $L_y$.}
    \label{fig:domain}
\end{figure}

\begin{figure}
    \centering
    \includegraphics[width=1\linewidth]{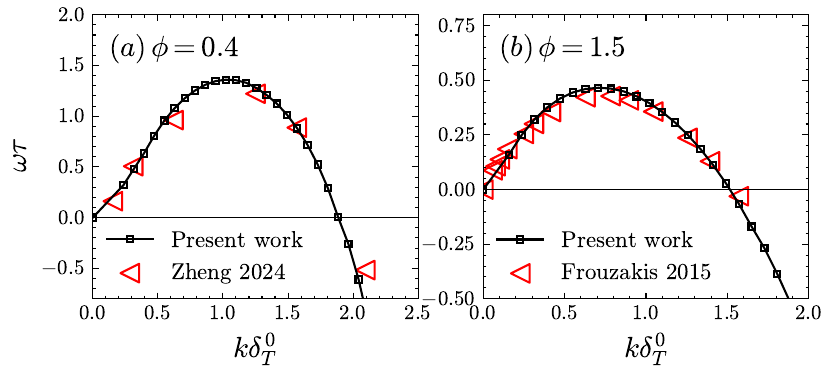}
    \caption{Comparison of the dispersion relations computed in the present work with previous results reported in the literature~\cite{zheng2024stability, frouzakis2015numerical}.}
    \label{fig:validation}
\end{figure}

\begin{figure*}[t]
    \centering
    \includegraphics[width=0.75\linewidth]{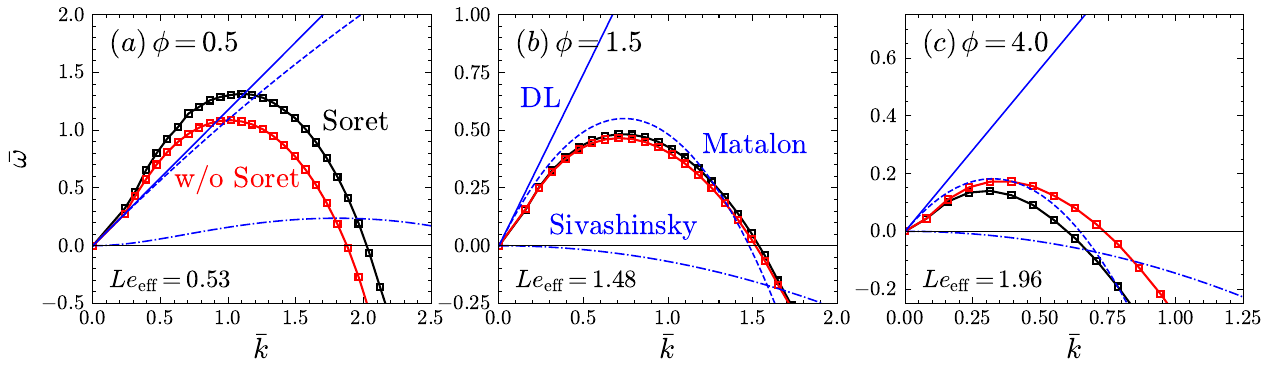}
    \caption{Dispersion relations at various equivalence ratios \(\phi\) between 0.5 (lean) and 4.0 (rich). Results with and without Soret diffusion are shown by black and red lines, respectively. The theoretical predictions are represented by blue lines, with solid, dashed, and dash-dotted styles corresponding to the DL~\cite{darrieus1938propagation,landau1988theory}, Matalon~\cite{matalon2003hydrodynamic}, and Sivashinsky~\cite{sivashinsky1977diffusional,Sivashinsky1983Instabilities} models, respectively. Also annotated in the figure are the effective Lewis numbers $Le_{\mathrm{eff}}$.}
    \label{fig:varies_phi_disp}
\end{figure*}

In this study, a series of direct numerical simulations (DNS) were conducted for hydrogen-air premixed flames in a two-dimensional rectangular computational domain. A schematic of the computation domain and boundary conditions is shown in Fig.~\ref{fig:domain}. The dimensions of the computation domain are normalized by the thermal thickness of a freely propagating one-dimensional unstretched flame, $\delta_T^0 = (T_b - T_u)/\max(|\nabla T|)$, where $S_L^0$, $T_u$ and $T_b$ denote the laminar flame speed, the unburnt gas temperature and the burnt/adiabatic flame temperature, respectively. For the linear stability analysis, the domain length and width are set to $L_x = 20\delta_T^0$ and $L_y = 80\delta_T^0$, respectively. Multi-wavenumber perturbations~\cite{al2024efficient} are imposed on the initial flame front to seed flame instabilities, with a perturbation amplitude of $A_0 = 1 \times 10^{-3}\delta_T^0$ to ensure a sufficiently long linear growth regime for accurate determination of the dispersion relations. A uniform grid with spatial resolution $\Delta x = \Delta y = \delta_T^0 / 30$ is employed. Results under standard conditions ($T_u = 300$~K, $P_u = 1$~atm) are discussed in the main text, while additional simulations at elevated temperatures and pressures are provided in the Supplementary Material.

The numerical method used in this work is validated by comparing the calculated dispersion relations with literature data~\cite{zheng2024stability, frouzakis2015numerical}, as demonstrated in Figure~\ref{fig:validation}. The wavenumber and growth rate are nondimensionalized as $\bar{k} = k\delta_T^0$ and $\bar{\omega} = \omega\tau$, respectively, with $\tau = \delta_T^0 / S_L^0$ denoting the characteristic flame time. Excellent agreement is observed under both lean and rich conditions, confirming the fidelity of the present simulations.

For simulating the nonlinear flame evolution at long times, a grid size of $\delta_T^0 /10$~\cite{zheng2024stability} is used to improve the computation efficiency. The domain length is extended to $L_x = 400\delta_T^0$ to capture the long-term evolution of the perturbed flame, while the domain width is increased to $L_y = 200\delta_T^0$ to avoid  confinement effects in the transverse direction~\cite{berger2019characteristic, zheng2024stability}. 

The present simulations are performed using the in-house code EBIdnsFOAM~\cite{zirwes2023assessment}. The reaction terms are modeled with finite-rate chemistry based on the detailed hydrogen-air reaction mechanism proposed by Li et al.~\cite{li2004updated}. A mixture-averaged transport model that includes Soret diffusion is employed~\cite{schlup2018validation, zirwes2025assessment}. Grid-independence and transport-model validations are presented in the Supplementary Material.

\section{Linear stability analysis} \label{sec:linear} \addvspace{10pt} 
\subsection{Linear growth rate}\label{sec:growth_rate}\addvspace{10pt}

Figure~\ref{fig:varies_phi_disp} presents the dispersion relations calculated for different equivalence ratios. For reference, the theoretical dispersion relations predicted by the Darrieus-Landau~\cite{darrieus1938propagation, landau1988theory}, Matalon~\cite{matalon2003hydrodynamic}, and Sivashinsky~\cite{sivashinsky1977diffusional,Sivashinsky1983Instabilities} models (see the Supplementary Material) are also shown in the figure; these models do not account for Soret diffusion. The DL model captures only the hydrodynamic effect induced by thermal expansion across the flame front, and its dispersion relation is given by~\cite{darrieus1938propagation, landau1988theory}:
\begin{equation}
    \bar\omega = \omega_{DL}\bar{k}= \frac{\sqrt{\sigma^3+\sigma^2-\sigma}-\sigma}{\sigma+1}\bar k,
\end{equation}
where $\sigma = \rho_u / \rho_b$ denotes the thermal expansion ratio, with $\rho_u$ and $\rho_b$ representing the densities of the unburnt and burnt gas, respectively. 

At \(\phi=0.5\), the TD instability is prominent. For small \(\bar{k}\), the dispersion curves lie above the DL line, indicating a strong destabilizing effect induced by differential diffusion. In addition, the instability is significantly enhanced when Soret diffusion is present -- both the maximum growth rate and the cutoff wavenumber are seen to increase. The Matalon model, which neglects the fourth-order stabilization term, overpredicts the growth rate at high wavenumbers. The Sivashinsky model, which assumes constant density and omits the DL instability, yields a growth rate significantly lower than the present simulations. At \(\phi=1.5\), the influence of Soret diffusion on the dispersion relation becomes negligible, as the growth rate changes little with and without the Soret effect. At \(\phi=4.0\), the role of Soret diffusion is reversed; it reduces the growth rate and exerts a stabilizing effect on the instability. For rich mixtures, the Matalon model provides better accuracy than the other models.

\begin{figure}
    \centering
    \includegraphics[width=1\linewidth]{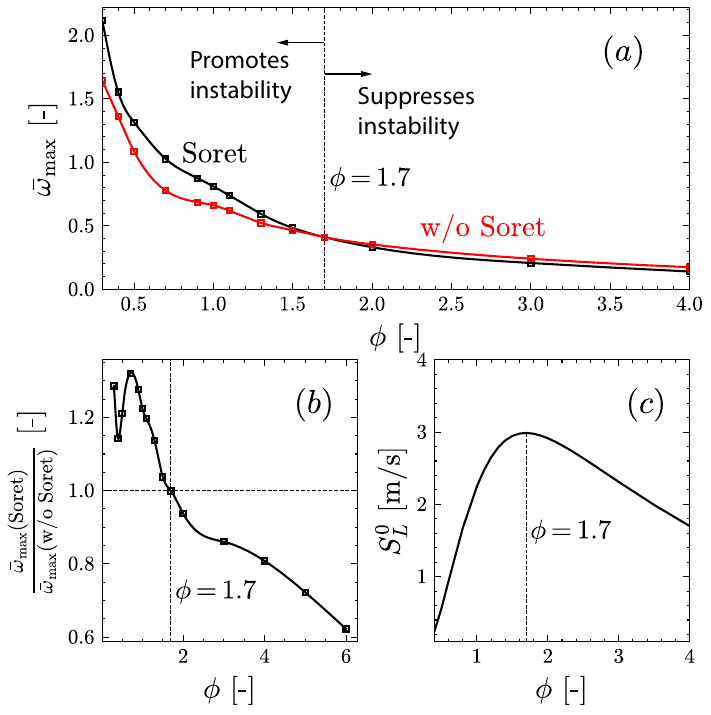}
    \caption{(a) Variation of the maximum growth rate $\bar{\omega}_{\mathrm{max}}$ with the equivalence ratio $\phi$, calculated with (black) and without (red) Soret diffusion. (b) The ratio between $\bar{\omega}_{\mathrm{max}}$ with and without Soret diffusion. (c) The unstretched laminar flame speed $S_L^0$ of \ce{H2}/air mixtures as a function of $\phi$, calculated without Soret diffusion. A critical point of $\phi=1.7$ is consistently observed in (b) and (c).}
    \label{fig:varies_phi_cha}
\end{figure}

Fig.~\ref{fig:varies_phi_cha}(a) shows the variation of the maximum nondimensional growth rate $\bar{\omega}_{\mathrm{max}}$ with the equivalence ratio $\phi$. Soret diffusion increases $\bar{\omega}_{\mathrm{max}}$ for lean and slightly rich mixtures, but its influence weakens as $\phi$ increases. Under very rich conditions, Soret diffusion reduces $\bar{\omega}_{\mathrm{max}}$, with a turning point at $\phi = 1.7$. The ratio of maximum growth rates with and without Soret diffusion is illustrated in Fig.~\ref{fig:varies_phi_cha}(b). This ratio generally decreases with increasing $\phi$, except for the range $0.4 \leq \phi \leq 0.9$. Under lean conditions, Soret diffusion can raise $\bar{\omega}_{\mathrm{max}}$ by up to 30\%, whereas on the very rich side it can suppress the instability growth rate by up to 40\%. 

Notably, the turning point of \(\phi=1.7\) is seen to coincide with the equivalence ratio at which the maximum laminar flame speed (in the absence of flame stretch and Soret diffusion) is achieved, as illustrated in Fig.~\ref{fig:varies_phi_cha}(c). The underlying physical mechanism will be discussed in Section 4.3. In addition, a non-monotonic behavior is observed under lean conditions, which may be related to a sign change in the Markstein length (the stretch sensitivity of flame speed) around $\phi = 0.6$ (without Soret diffusion) or $\phi = 0.7$ (with Soret diffusion). The general effects of Soret diffusion on the intrinsic flame instability are qualitatively consistent across different initial temperatures and pressures (see Supplementary Material).

\subsection{Markstein length}\label{sec:Markstein}\addvspace{10pt}

Additional analysis about the effect of Soret diffusion on the Markstein length was conducted under a numerical configuration of one-dimensional premixed counterflow twin flames. The simulations were performed with Cantera~\cite{goodwin2018cantera} using 
the same detailed kinetic mechanism and mixture-averaged transport model as used in the 2D DNS. The flame strain rate $K_s$ was obtained from the maximum velocity gradient just upstream of the flame and was varied by adjusting the unburnt gas velocity at the opposing nozzles. The Markstein length $\mathcal{L}$, together with the unstretched laminar flame speed $S_L^0$, was determined from the linear relationship between the stretched laminar flame speed $S_L$ and $K_s$ in the low-stretch regime:

\begin{equation}
    S_L = S_L^0 - \mathcal{L} \cdot K_s
\end{equation}

\begin{figure}
    \centering
    \includegraphics[width=1\linewidth]{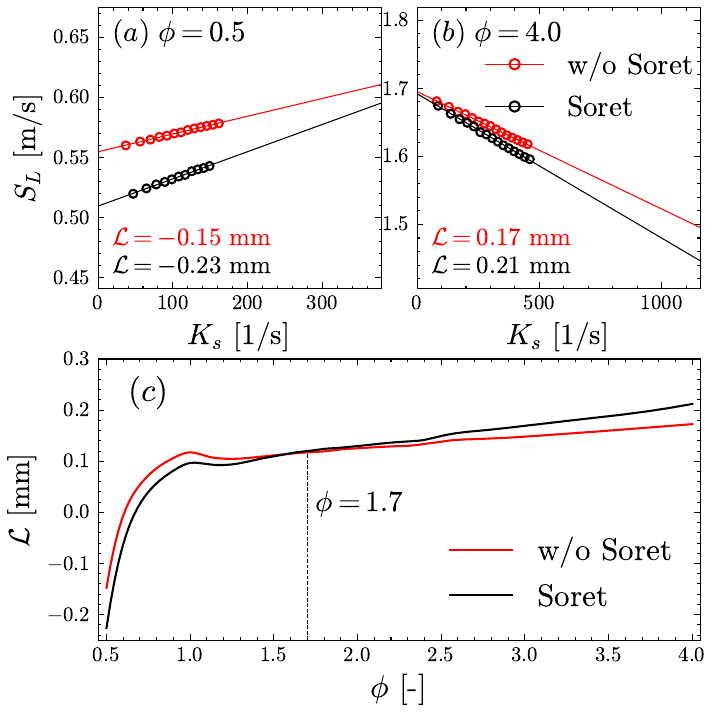}
    \caption{The effect of Soret diffusion on the Markstein length. (a)-(b) Variation of the stretched laminar flame speed $S_L$ for \ce{H2}/air mixtures with the stretch rate $K_s$ under lean ($\phi = 0.5$) and rich ($\phi = 4.0$) conditions. Symbols represent direct simulation results, while lines represent linear fits in the low-stretch regime. The Markstein lengths $\mathcal{L}$, given by the negative slopes of the fits, are displayed in the panels. (c) Equivalence-ratio dependence of the Markstein length, calculated with (black) and without (red) Soret diffusion.}
    \label{fig:Lb}
\end{figure}

The determination of $\mathcal{L}$ is illustrated in Fig.~\ref{fig:Lb}(a) and (b) for lean ($\phi=0.5$) and rich ($\phi=4.0$) mixtures, respectively. At $\phi=0.5$, $S_L$ increases with the stretch rate $K_s$, and the inclusion of Soret diffusion leads to a steeper slope, corresponding to a more negative $\mathcal{L}$. Conversely, at $\phi=4.0$, Soret diffusion leads to a more positive $\mathcal{L}$.

Fig.~\ref{fig:Lb}(c) shows the variation of $\mathcal{L}$ over a broader range of $\phi$. The results indicate that Soret diffusion reduces $\mathcal{L}$ for $\phi < 1.7$ and increases $\mathcal{L}$ for $\phi > 1.7$. Since a smaller Markstein length corresponds to stronger stretch-induced instability, these 1D counterflow results corroborate the 2D DNS observation: Soret diffusion enhances instability in lean to slightly rich mixtures and suppresses it under very rich conditions.

\section{Nonlinear flame evolution}\label{sec:non_linear}\addvspace{10pt}

\subsection{Multi-scale flame morphology} \label{sec:morphology}\addvspace{10pt}

\begin{figure}
    \centering
    \includegraphics[width=0.75\linewidth]{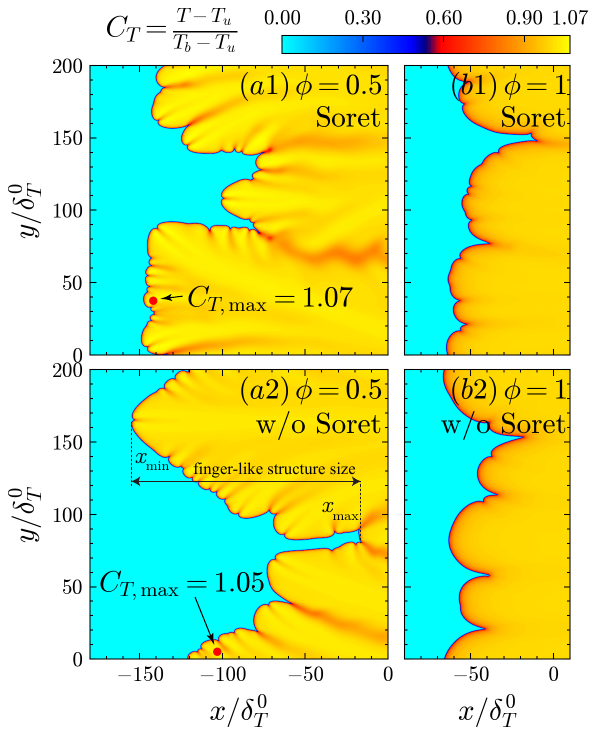}
    \caption{Dimensionless temperature contours of fully developed flames at different equivalence ratios. Results in the top two and the bottom two panels are calculated with and without Soret diffusion, respectively.}
    \label{fig:contour}
\end{figure}

The influence of Soret diffusion on flame morphology in the nonlinear regime is examined under lean and stoichiometric conditions, where the flame fronts are highly irregular and exhibit cellular structures of various scales. Representative results are shown in 
Fig.~\ref{fig:contour}, where the nondimensional temperature, $C_T = (T - T_u) / (T_b - T_u)$, is used as the progress variable for visualization. Under lean conditions, the TD instability, driven by differential diffusion between species and heat, yields super-adiabatic temperature distributions ($C_{T,\mathrm{max}}>1$), a tendency that is further amplified by Soret diffusion. For example, at $\phi$ = 0.5, $C_{T,\mathrm{max}}$ increases from 1.05 to 1.07 when Soret diffusion is included.

To systematically characterize the flame-front morphology, two distinct spatial scales are defined. The macro-scale metric ($x_{\max} - x_{\min}$, Fig.~\ref{fig:contour}a2) quantifies the streamwise extent of the flame. It essentially measures the elongated finger size under lean conditions (or the DL-driven cusp depth under stoichiometric conditions). The second scale captures small-scale cellular structures, with characteristic sizes denoted as $\lambda_\text{cell}$, defined by the arc length between two adjacent curvature minima along the flame front~\cite{berger2019characteristic}. The flame front is identified as the isoline of the product mass fraction, $Y_{\mathrm{H_2O}} = Y^*_{\mathrm{H_2O}}$, where $Y^*_{\mathrm{H_2O}}$ is the $\mathrm{H_2O}$ mass fraction at the location of maximum heat release rate in a planar reference flame of the same equivalence ratio. 

\begin{figure}
    \centering
    \includegraphics[width=1\linewidth]{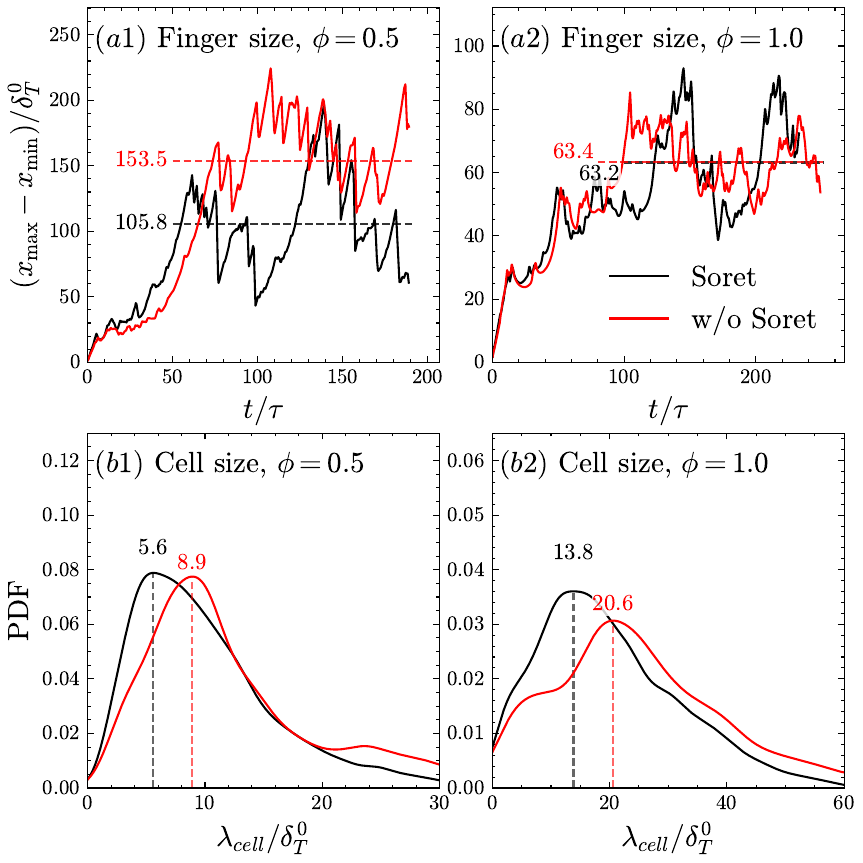}
    \caption{(a) Temporal evolution of the finger size with (black) and without (red) Soret diffusion. Dashed lines indicate time-averaged values. (b) Probability density functions of the cell size $\lambda_\text{cell}$. The left and right panels correspond to lean and stoichiometric flames, respectively.}
    \label{fig:cell_size}
\end{figure}

As shown in Fig.~\ref{fig:contour}(a2), large finger-like structures appear on the flame front under lean conditions; however, their sizes are substantially reduced when Soret diffusion is included~(see Fig.~\ref{fig:contour}(a1)). The temporal evolution of the finger size, with and without Soret diffusion, is depicted in Fig.~\ref{fig:cell_size}(a). Quantitatively, the time-averaged finger size, as shown in Fig.~\ref{fig:cell_size}(a1),  decreases by approximately one-third with Soret diffusion. Meanwhile, the most probable cell size is reduced by approximately 40\% (see Fig.~\ref{fig:cell_size}(b1)). For lean hydrogen flames, the Soret effect accelerates the formation and splitting of small-scale wrinkles and hinders their subsequent evolution into large-scale structures by amplifying the influence of preferential diffusion.

Under stoichiometric conditions ($\phi$ = 1.0), the TD instability is weak, and the development of finger-like structures is dominated by the DL instability. Consequently, as shown in Fig.~\ref{fig:cell_size}(a2), the influence of Soret diffusion on the finger size is marginal. However, Soret diffusion significantly reduces the characteristic size of small-scale cells (Fig.~\ref{fig:cell_size}(b2)). By promoting \ce{H2} transport and enhancing preferential diffusion, Soret diffusion intensifies the TD instability and increases the amount of small-scale wrinkles. This morphological shift is evident when comparing flame contours in Fig.~\ref{fig:contour}(b1) and Fig.~\ref{fig:contour}(b2).

\subsection{Flame speed analysis} \label{sec:flame_speed} \addvspace{10pt}

The present study also evaluates the influence of Soret diffusion on the fuel consumption speed $S_c$, an integral quantity representing the domain-averaged fuel depletion rate. $S_c$ is defined as:
\begin{equation}
S_c = -\frac{1}{\rho_u L_y (Y_{\mathrm{H_2},u}-Y_{\mathrm{H_2},b})}\iint_{\Omega}\dot \omega_\mathrm{H_2} \mathrm{d}x\mathrm{d}y.
\end{equation}
In this equation, $Y_{\mathrm{H_2},u}$ and $Y_{\mathrm{H_2},b}$ are the mass fractions of H$_2$ in the unburnt and burnt gas, respectively, and $\dot{\omega}_\mathrm{H_2}$ is the mass consumption rate of hydrogen. 

The flame acceleration is governed by two mechanisms \cite{berger2019characteristic}: (1) flame wrinkling that increases the total flame surface area $A$ (which, in 2D simulations, reduces to the arc length of the flame front) above the reference value $L_y$; and (2) the enhancement of local propagation speed due to flame stretch, quantified by the stretch factor $I$. Here, $A$ is computed using the generalized flame surface density (GFSD) formalism \cite{vervisch1995surface}, with the progress variable defined as $Y_{\mathrm{H_2O}}$ for rich and stoichiometric flames and $Y_{\mathrm{H_2}}$ for lean flames to avoid overestimation of the area from post-flame trails induced by strong TD instability ~\cite{berger2022intrinsic_b, howarth2022empirical}. The combined effect of these two mechanisms can be expressed as~\cite{berger2019characteristic}:

\begin{equation}
    \frac{S_c}{S_L^0} = \frac{A}{L_y}I
\end{equation}

\begin{figure}
    \centering
    \includegraphics[width=1\linewidth]{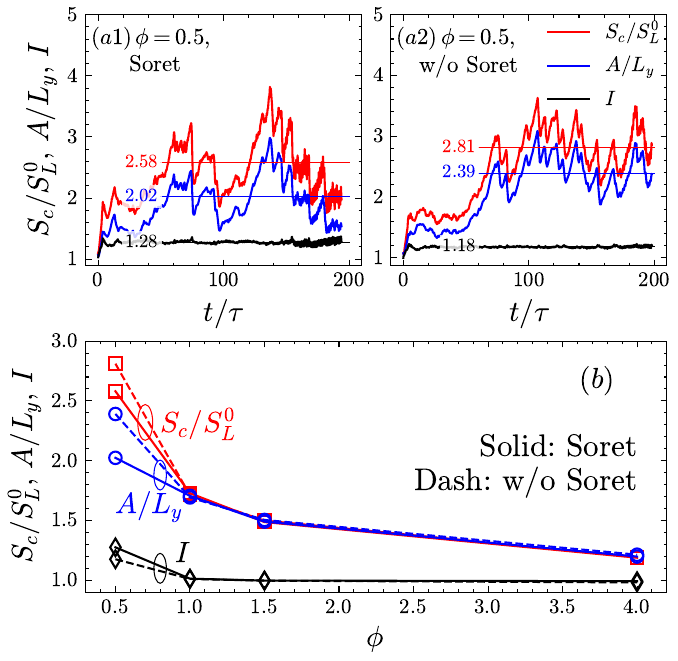}
    \caption{Temporal evolution of the normalized fuel consumption speed $S_c/S_L^0$, the flame surface area ratio $A/L_y$, and the stretch factor $I$ at $\phi = 0.5$ with (a1) and without (a2) Soret diffusion. Horizontal lines indicate time-averaged values. (b) Time-averaged $S_c/S_L^0$, $A/L_y$, and $I$ as functions of $\phi$.}
    \label{fig:flame_speed_res}
\end{figure}

The time-histories and time-averaged values of $S_c/S_L^0$, $A/L_y$ and $I$ for $\phi=0.5$, with and without the Soret effect, are presented in Fig.~\ref{fig:flame_speed_res}. A counter-intuitive observation is that, although Soret diffusion enhances flame instability under lean conditions, it reduces the normalized consumption speed $S_c/S_L^0$. Further analysis of the contributing factors reveals that the stretch factor $I$ is moderately increased by Soret diffusion, but the flame surface area ratio $A/L_y$ is significantly reduced and dominates the reduction in $S_c/S_L^0$. This aligns with the observed decrease in finger size, which reduces the flame surface area. The temporal fluctuations of $S_c/S_L^0$ are more pronounced in the presence of Soret diffusion, suggesting stronger flow pulsations driven by flame instability.

Figure~\ref{fig:flame_speed_res}(b) shows the variation of $S_c/S_L^0$, $A/L_y$ and $I$ with $\phi$. The effect of Soret diffusion on $S_c/S_L^0$ is most pronounced in the lean regime but becomes negligible at stoichiometric and rich conditions. Both $S_c/S_L^0$ and $A/L_y$ decrease monotonically as $\phi$ increases, while the stretch factor $I$ remains near unity for $\phi \geq 1$. 

\begin{figure}
    \centering
    \includegraphics[width=1\linewidth]{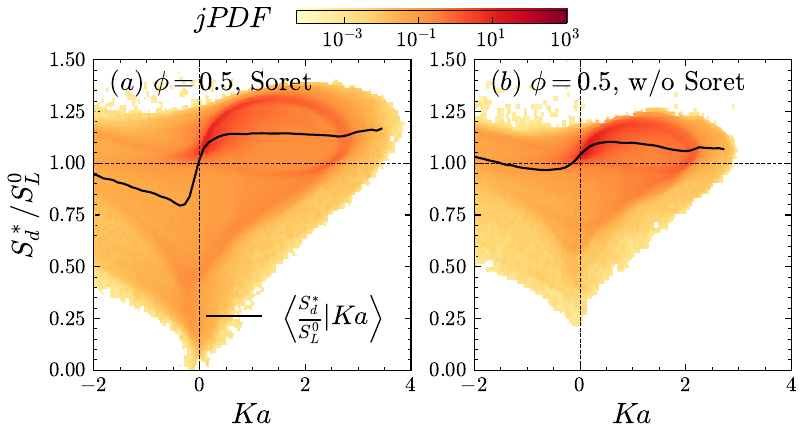}
    \caption{Joint probability density function (dot clouds) of the normalized density-weighted displacement speed $S_d^*/S_L^0$ and the Karlovitz number $Ka$, shown together with the conditional mean of $S_d^*/S_L^0$ given $Ka$ (black lines). Results are presented for $\phi = 0.5$ with (a) and without (b) Soret diffusion.}   
    \label{fig:Sd_Ka_relation}
\end{figure}

Additional analysis investigates the effect of Soret diffusion on the flame displacement speed. The local flame-stretch interaction is quantified using the density-weighted displacement speed $S_d^*=\rho S_d/\rho_u$ and the Karlovitz number $Ka=K_s\tau$ (detailed formulations are provided in the Supplementary Material). All quantities are evaluated at the flame front.

As shown in Fig.~\ref{fig:Sd_Ka_relation}, the joint probability density function (jPDF) of $S_d^*/S_L^0$ and $Ka$ is strongly affected by Soret diffusion, particularly away from the distribution center. The broad jPDF indicates that local reactivity is modulated along the highly wrinkled cellular flame front. Additionally, the conditional mean $\langle S_d^*/S_L^0 |Ka \rangle$ exhibits a positive correlation with the stretch rate at small $|Ka|$, which is consistent with a negative Markstein length. However, this correlation no longer applies at large $|Ka|$, where the competition between flame stretch and heat release via radical recombination becomes dominant \cite{sanchez2014recent}. The stretch sensitivity is amplified by Soret diffusion; under positive/negative stretch, the local flame speed is substantially higher/lower with Soret diffusion, thereby intensifying TD instability in lean mixtures.

\begin{figure}
    \centering
    \includegraphics[width=0.75\linewidth]{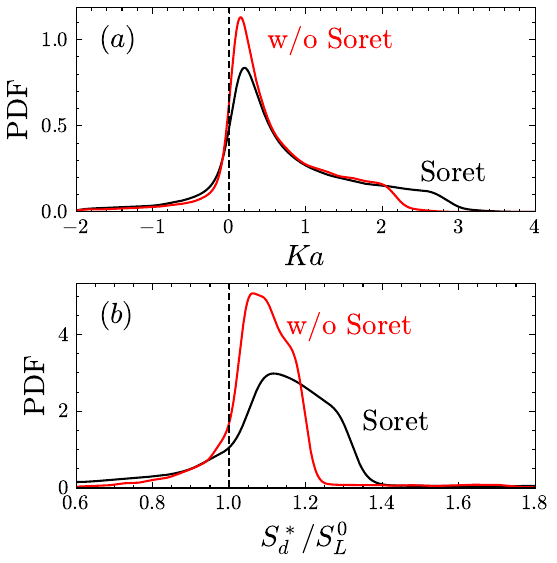}
    \caption{Probability density functions of the Karlovitz number $Ka$ (a) and the normalized density-weighted displacement speed $S_d^*/S_L^0$ (b) for $\phi = 0.5$, with (black) and without (red) Soret diffusion.}
    \label{fig:Sd_Ka_pdf}
\end{figure}

Fig.~\ref{fig:Sd_Ka_pdf}(a) presents the PDF of the Karlovitz number $Ka$. As discussed previously, Soret diffusion reduces the size of cellular structures~(Fig.~\ref{fig:cell_size}(b1)), leading to smaller radii of curvature and consequently higher local stretch rates along the highly wrinkled flame front. Soret diffusion slightly reduces the probability of low stretch rates ($Ka < 0.5$) while significantly increasing the probability of high stretch rates (e.g., $Ka > 2$). As a result, the distribution becomes broader and extends further toward the positive side.

Fig.~\ref{fig:Sd_Ka_pdf}(b) shows the PDF of $S_d^*/S_L^0$. The PDF of $S_d^*/S_L^0$ also exhibits a noticeable rightward shift when Soret diffusion is considered, which is consistent with the previously mentioned positive correlation between $S_d^*$ and $Ka$. The probability of significant enhancement of the local displacement speed ($S_d^*/S_L^0 > 1.25$) increases considerably. From a statistical perspective, it indicates that Soret diffusion effectively increases the local flame displacement speed by amplifying preferential diffusion in highly stretched regions.

\subsection{Flame segment analysis} \label{sec:FSA}\addvspace{10pt}

\begin{figure}
    \centering
    \includegraphics[width=0.75\linewidth]{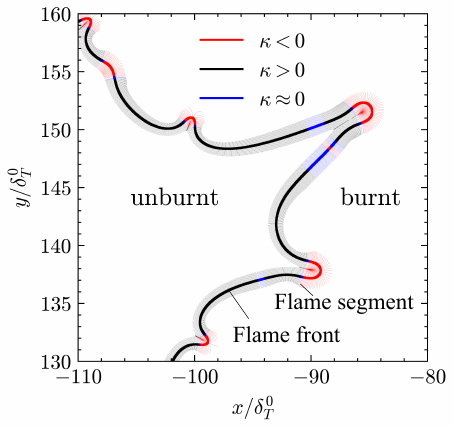}
    \caption{Partition of the flame front into three regions based on the local curvature $\kappa$. Each flame segment (shaded band) extends from the flame front (thick line) toward both the unburnt and burnt sides.}
    \label{fig:segment_region}
\end{figure}

The effect of Soret diffusion on transport characteristics in regions of different curvatures is investigated using a flame segment analysis method adapted from previous studies~\cite{day2009turbulence, wen2023numerical, lehmann2025effects}. Each segment extends into the pre- and post-flame regions along the gradient of the progress variable $\nabla C_{\mathrm{H_2O}}$ (where $C_\mathrm{H_2O}=Y_\mathrm{H_2O}/Y_{\mathrm{H_2O},b}$, with $Y_{\mathrm{H_2O},b}$ being the \ce{H2O} mass fraction in the burnt gas), ensuring a complete representation of the local flame structure . The flame segments are categorized by the centerline curvature \(\kappa\), with the positively and negatively curved regions defined as \(\kappa>0.01\kappa_{\mathrm{max}}\) and \(\kappa<0.01\kappa_{\mathrm{min}}\), respectively~\cite{lehmann2025effects}. A conceptual illustration of the partition of flame segments is shown in Fig.~\ref{fig:segment_region}. 

\begin{figure*}
    \centering
    \includegraphics[width=0.75\linewidth]{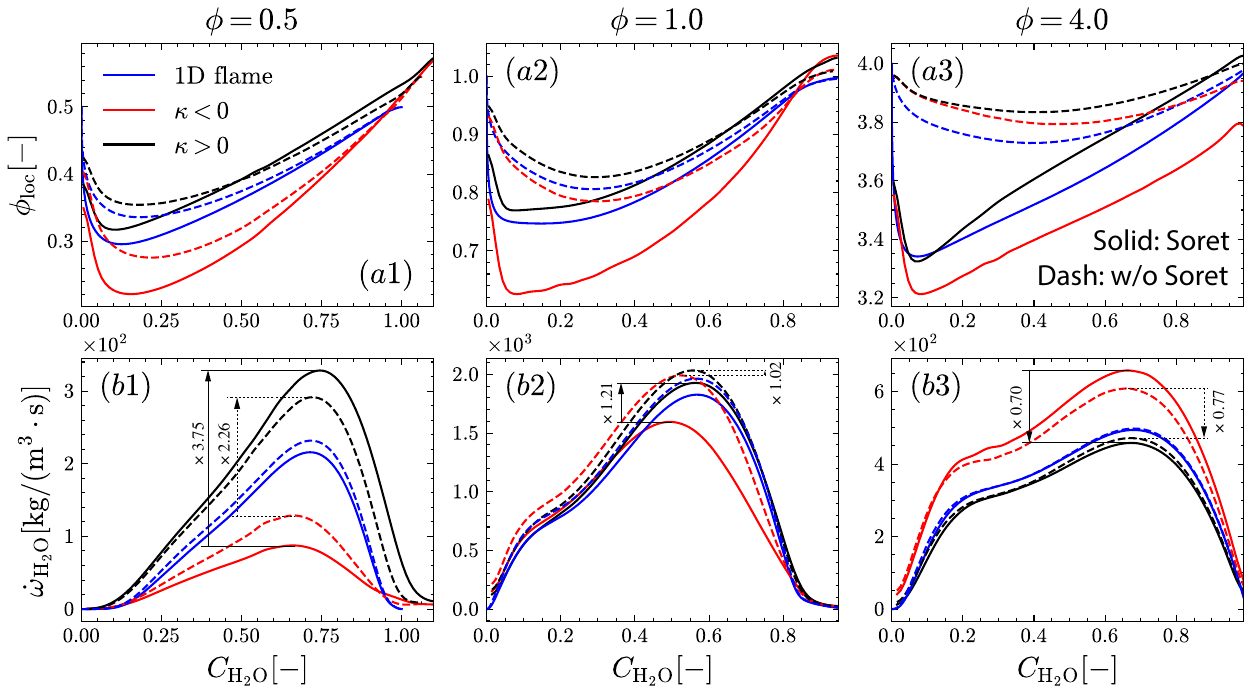}
    \caption{Distributions of the local equivalence ratio $\phi_{\mathrm{loc}}$ (a) and the mass production rate $\dot{\omega}_{\mathrm{H_2O}}$ (b) along flame segments of different curvature types. Results are shown for lean ($\phi=0.5$), stoichiometric ($\phi=1.0$), and rich ($\phi=4.0$) flames with (solid lines) and without (dashed lines) Soret diffusion. The bottom panels also display the ratios of the peak $\dot{\omega}_{\mathrm{H_2O}}$ between the positively and negatively curved regions.}
    \label{fig:FSA}
\end{figure*}

Fig.~\ref{fig:FSA} shows the mean values of the local equivalence ratio (\(\phi_{\mathrm{loc}} = 8\xi_\mathrm{H}/\xi_\mathrm{O}\), where \(\xi_\mathrm{H}\) and \(\xi_\mathrm{O}\) denote the local mass fractions of elements H and O, respectively; see the Supplementary Material) and the mass production rate of water (\(\dot{\omega}_{\mathrm{H_2O}}\)) in regions of different flame curvatures as functions of the progress variable \(C_\mathrm{H_2O}\). Across all conditions explored in the present study, the influence of Soret diffusion on \(\phi_\mathrm{loc}\), along with its underlying physical mechanism, is consistent. When Soret diffusion is neglected and only Fickian diffusion is 2considered, the large diffusion coefficient of hydrogen molecules causes the fuel to accumulate in convex regions (\(\kappa>0\)) and to dissipate in concave regions (\(\kappa<0\)). This explains the dominant trend observed in the top panels of Fig.~\ref{fig:FSA}: for all cases, the local equivalence ratio $\phi_{\mathrm{loc}}$ in the convex regions is consistently higher than that in the concave regions, with both deviating in opposite directions from the reference value of planar flame. When Soret diffusion is further considered, the difference in \(\phi_{\mathrm{loc}}\) between the two curvature regions is amplified. This occurs because, near the flame front, the temperature gradient \(\nabla T\) and the concentration gradient \(\nabla Y_{\mathrm{H_2}}\) typically align, and Soret diffusion drives \ce{H2} transport in the same direction as Fickian diffusion.

The influence of Soret diffusion on the reaction rate, unlike that on the local equivalence ratio, is highly dependent on the global mixture conditions. For the lean case at \(\phi = 0.5\), the reaction rate is mainly governed by the local \ce{H_2} concentration. Preferential diffusion leads to \(\mathrm{H_2}\) enrichment in the positively curved region, where the peak \(\dot{\omega}_{\mathrm{H_2O}}\) is 2.26 ($>1$) times that in the negatively curved region. Consequently, the local flame speeds of convex segments are accelerated relative to those of concave segments, causing the flame front to protrude further into the unburnt mixture. This positive feedback is responsible for the continuous development of wrinkles and represents the essence of TD instability. When Soret diffusion is included, this ratio increases to 3.75, leading to a strong enhancement of TD instability. 

In contrast, for the rich case at \(\phi = 4.0\), the reaction rate is dominated by the local \ce{O_2} concentration, and fuel enrichment reduces the reaction rate in the positively curved region. The corresponding peak \(\dot{\omega}_{\mathrm{H_2O}}\) is 0.77 ($<1$) times that in the negatively curved region, suggesting that the flame is TD-stable. When Soret diffusion is considered, the ratio decreases to 0.70, and the stabilization effect becomes more pronounced.

For the stoichiometric flame, the preferential diffusion effect is weak in the absence of Soret diffusion, and the relative difference between peak $\dot{\omega}_{\mathrm{H_2O}}$ in the positively and negatively curved regions is less than 2\%. However, Soret diffusion increases this difference to 21\% via two distinct mechanisms. On the one hand, temperature gradients across curved flame fronts drive local \ce{H2} accumulation/dissipation in a manner similar to preferential diffusion, thereby modifying the local reaction rate according to the sign of curvature. On the other hand, the enhanced transport of \ce{H} radicals toward the burnt gas reduces the overall reaction rate~\cite{liang2013effects}, causing the peak $\dot{\omega}_{\mathrm{H_2O}}$ to decrease for all curvatures. The overall effect is that the flame becomes apparently TD-unstable.

The local enrichment/dilution effect induced by Soret diffusion can also explain the aforementioned sensitivity reversal (see Fig.~\ref{fig:varies_phi_cha}) near the critical equivalence ratio ($\phi_c$ = 1.7) corresponding to the maximum unstretched laminar flame speed. Soret diffusion always increases \(\phi_{\mathrm{loc}}\) in the positively curved region but has opposite effects on the local flame speed under conditions of \(\phi < 1.7\) and \(\phi > 1.7\), thereby destabilizing and stabilizing the flame, respectively.

\section{Conclusions}\label{sec:con}\addvspace{10pt}

The role of Soret diffusion in the instability evolution dynamics of premixed hydrogen flames is quantitatively analyzed under standard temperature and pressure. In the linear regime, Soret diffusion increases the perturbation growth rate for $\phi < 1.7$, especially under lean conditions, but reduce the growth rate at $\phi > 1.7$. The turning point of $\phi_c = 1.7$ coincides with the equivalence ratio corresponding to the maximum unstretched laminar flame speed, where a similar sensitivity reversal is also observed in the Markstein length ($\mathcal{L}$). In the nonlinear regime, Soret diffusion significantly modifies the morphology of lean flames by accelerating the formation of small-scale wrinkles and reducing the characteristic size of finger-like structures. However, although Soret diffusion promotes preferential diffusion and increases the local flame displacement speed under lean conditions, it significantly reduces the overall flame surface area and consequently the global fuel consumption rate. 

The probability distributions of the Karlovitz number ($Ka$) and the density-weighted displacement speed ($S^*_d$) are also analyzed. The results suggest that, for lean hydrogen flames, Soret diffusion broadens the probability distributions of both parameters, especially on the positive side. Further analysis of flame segments reveals that Soret diffusion acts synergistically with Fickian diffusion to modify the local equivalence ratio $\phi_{\mathrm{loc}}$ according to the local curvature, which explains the sensitivity reversal at $\phi_c$ = 1.7. Notably, the qualitative trend of Soret diffusion enhancing instability in lean mixtures and suppressing it in highly rich mixtures is generally valid across a wide range of temperatures and pressures. However, the quantitative results from nonlinear analysis -- such as the numerical metrics of flame morphology, global consumption speeds, and the exact values of critical equivalence ratio ($\phi_c$)-- depend on the specific flame conditions. This study contributes to the understanding of hydrogen flame dynamics under complex differential transport across an extended range of equivalence ratios and through different regimes of instability evolution.

\acknowledgement{CRediT authorship contribution statement} \addvspace{10pt}

{\bf Q.~W.}: Investigation; Formal analysis; Visualization; Writing -- original draft. {\bf Y.~W.}: Formal analysis; Visualization; Writing -- review \& editing. {\bf L.~Y.}: Methodology; Validation; Writing -- review \& editing. {\bf Y.~M.}: Supervision; Project administration; Writing -- review \& editing. {\bf T.~Z.}: Software; Writing -- review \& editing. {\bf S.~W.}: Conceptualization; Methodology; Supervision; Writing -- review \& editing. {\bf Z.~C.}: Conceptualization; Resource; Supervision; Writing -- review \& editing. {\bf All authors}: Discussed the results and approved the final manuscript.

\acknowledgement{Declaration of competing interest}
\addvspace{10pt}

The authors declare that they have no known competing financial interests or personal relationships that could have appeared to influence the work reported in this paper.

\acknowledgement{Acknowledgments}
\addvspace{10pt}

This work was supported by the National Natural Science Foundation of China under Grants No. 52425604 and by the National Key Research and Development Program of China under Grant No. 2025YFF0511801.

\footnotesize
\baselineskip 9pt

\clearpage
\thispagestyle{empty}
\bibliographystyle{proci}
\bibliography{PROCI_LaTeX}


\newpage

\small
\baselineskip 10pt


\end{document}